\documentclass[12pt]{article}

\usepackage{latexsym,amsfonts,amsmath,amsthm,amssymb}

\usepackage[T1]{fontenc}
\usepackage[latin1]{inputenc}
\title{On the classical limit of hyperbolic quantum mechanics}
\author{Andrei Khrennikov\\
School of Mathematics and Systems Engineering\\
University of V\"axj\"o, S-35195, Sweden}
\begin{document}
\maketitle

\abstract{We demonstrated that classical mechanics have, besides the well known
quantum deformation, another deformation -- so called hyperbolic quantum
mechanics. The classical Poisson bracket can be obtained as the limit $h\to 0$
not only of the ordinary Moyal bracket, but also hyperbolic analogue of the
Moyal bracket. Thus there are two different deformations of classical phase-space:
complex Hilbert space and hyperbolic Hilbert space (module over a so called
hyperbolic algebra -- the two dimensional Clifford algebra).
Ordinary (complex) and hyperbolic quantum mechanics are characterized
by two types of interference perturbation of the classical formula of total probability:
ordinary $\cos$-interference and hyperbolic $\cosh$-interference.}

\section{Introduction}
In a series of papers on probabilistic foundations of quantum theory [1] it was demonstrated that
depending on the mutual disturbance of observables the conventional formula of total probability
can be deformed in two ways. One deformation of the formula of total probability is the well known
formula for {\it{quantum interference of probabilities:}}
\begin{equation}
\label{I}
{\bf P}(b=b_j)={\bf P}(b=b_j/a=a_1) {\bf P}(a=a_1) + {\bf P}(b=b_j/a=a_2) {\bf P}(a=a_2)
\end{equation}
\[+ 2 \cos \theta \sqrt{{\bf P}(b=b_j/a=a_1) {\bf P}(a=a_1) {\bf P}(b=b_j/a=a_2) {\bf P}(a=a_2)}\]
Here $b=b_1, b_2$ and $a=a_1, a_2$ are two dichotomous random variables.
Thus the conventional formula of total probability is perturbed by the trigonometric term.
Such a ``formula of total probability with interference term''
can be easily derived by using the conventional quantum formalism: one should
make transition from the orthonormal basis $\{e_j^b\}_{j=1}^2$ (corresponding to the observable
$\hat b$) to the orthonormal basis $\{e_j^a\}_{j=1}^2$
(corresponding to the observable $\hat a$). We pay attention that in [1] this formula was derived
in purely classical (but contextual) probabilistic framework without to appeal to the
quantum formalism. Of course, a classical probabilistic reconstruction of noncommutative structure of QM is
important for better understanding of QM. However, the contextual probabilistic approach
does not only reproduce the well known quantum noncommutative structure. There appears naturally
another quantum-like structure which we called hyperbolic QM, see [1]. The starting point was that the contextual probabilistic formalism induces not only the conventional
quantum formula of total probability (\ref{I}), but also
another formula, see [1]:
\begin{equation}
\label{I1}
{\bf P}(b=b_j)={\bf P}(b=b_j/a=a_1) {\bf P}(a=a_1) + {\bf P}(b=b_j/a=a_2) {\bf P}(a=a_2)
\end{equation}
\[\pm 2  \cosh \theta \sqrt{{\bf P}(b=b_j/a=a_1) {\bf P}(a=a_1) {\bf P}(b=b_j/a=a_2) {\bf P}(a=a_2)}\]
(here $\theta$ cannot be arbitrary and the range of variation of $\theta$ depends on probability distributions
of observables $a$ and $b$).
In [1] it was demonstrated that the formula of total probability with the hyperbolic interference
term (\ref{I1}) also can be derived in the framework of the so called hyperbolic quantum mechanics.
In the hyperbolic
quantum mechanics [1] observables are represented by self-adjoint operators in a
``hyperbolic Hilbert space'' --  Hilbert module over the two-dimensional Clifford algebra $\bf G$ --
{\it hyperbolic algebra}.

It is interesting that hyperbolic
quantization also appeared naturally in relativistic quantum physics.
The hyperbolic numbers offer the possibility to represent
the four-component Dirac spinor as a two component hyperbolic spinor. Hucks has shown [2] that the
Lorentz group is equivalent to the hyperbolic unitary group. Poteous [3] proved the unitarity of special
linear group with the help of the double field, which corresponds to the null basis representation
of the hyperbolic numbers. S. Ulrych investigated the hyperbolic representation of
Poincare mass [4]. He also studied symmetries in the hyperbolic Hilbert space [4]. Applications of hyperbolic
 numbers in general relativity can be found in the paper [5] of G. Kunstatter et al. These intensive applications of hyperbolic numbers
 in quantum physics induces a natural question: What is a classical limit of the hyperbolic QM?

We study this problem in this note. We recall
that in the conventional quantum mechanics this problem is solved by using the {\it deformation quantization} framework,
see particularly the work of Moyal [6] (and, e.g., [7], [8] for the modern presentation
and various generalizations). In this framework it
is proved that when $h\to 0$ the Moyal bracket on the space $E(Q \times  P)$ of symbols
$a(q,p)$ of pseudo-differential operators (representing quantum observables) is
transformed into the Poisson bracket.

We use the same approach in the hyperbolic case. We develop a calculus of pseudo-differential operators over
the hyperbolic algebra $\bf G$ and found the limit of the hyperbolic Moyal bracket. Surprisingly we obtain
the standard Poisson bracket. Thus:

{\bf {The classical limit of hyperbolic quantum mechanics is ordinary
classical mechanics.}}

This derivation needs quite a lot of mathematics (which is similar to used in so called
functional superanalysis, see, e.g.   [9], [10], [8]). These mathematical investigations can be omitted and
physicists can go directly to theorems 2 and 3 and conclusions at the end of sections 2 and 3:

Classical mechanics can be deformed in two ways by using complex and hyperbolic representations.
These deformations describe two different types of interference of probabilities: the trigonometric
interference and the hyperbolic interference.

\section{Hyperbolic numbers}

  We introduce an involution in {\bf{G}} by setting
$\bar{z} = x - j y$
and set  $|z|^2=z\bar{z}=x^2-y^2.$
We remark that  $|z|=\sqrt{x^2-y^2}$ is not well defined for an arbitrary $z\in {{\bf{G}}}.$
We set ${{\bf{G}}}_+=
\{z\in{{\bf{G}}}:|z|^2\geq 0\}.$ We remark that ${{\bf{G}}}_+$
is a multiplicative semigroup as follows from the equality

$|z_1 z_2|^2=|z_1|^2 |z_2|^2.$

Thus, for $z_1, z_2 \in {{\bf{G}}}_+,$
we have $|z_1 z_2|=|z_1||z_2|.$ We introduce

$e^{j\theta}=\cosh\theta+ j \sinh\theta, \; \theta \in {\bf{R}}.$

We remark that

$e^{j\theta_1} e^{j\theta_2}=e^{j(\theta_1+\theta_2)}, \overline{e^{j\theta}}
=e^{-j\theta}, |e^{j\theta}|^2= \cosh^2\theta - \sinh^2\theta=1.$

Hence, $z=\pm e^{j\theta}$ always belongs to ${{\bf{G}}}_+.$
We also have

$\cosh\theta=\frac{e^{j\theta}+e^{-j\theta}}{2}, \;\;\sinh\theta=\frac{e^{j\theta}-e^{-j\theta}}{2 j}\;.$

We set ${{\bf{G}}}_+^*=
\{z\in{{\bf{G}}}_+:|z|^2>0 \}. $
Let  $z\in {{\bf{G}}}_+^*.$  We have

$z=|z|(\frac{x}{|z|}+j \frac{y}{|z|})= \rm{sign}\; x\; |z|\;(\frac{x {\rm{sign}} x}{|z|} +j\;
\frac{y {\rm{sign}} x}{|z|}).$

As $\frac{x^2}{|z|^2}-\frac{y^2}{|z|^2}=1,$  we can represent $x$ sign $x= \cosh\theta$
and $y$ sign $x=\sinh\theta, $ where the phase $\theta$ is unequally defined.
We can represent each $z\in {{\bf{G}}}_+^*$ as

$z = \rm{sign}\; x\;  |z|\; e^{j\theta}\;.$

By using this representation we can easily prove that ${{\bf{G}}}_+^*$
is a multiplicative group. Here $\frac{1}{z}=\frac{{\rm{sign}} x}{|z|}e^{-j\theta}.$
The unit circle in ${{\bf{G}}}$ is defined as $S_1 = \{z\in{{\bf{G}}}:|z|^2=1\}
=\{ z= \pm e^{j \theta}, \theta \in (-\infty, +\infty)\}.$ It is a multiplicative
subgroup of ${\bf G}_+^*.$

We remark that for any $y \in {\bf R}$ the map:
$$
{\bf R} \to {\bf G}, x\to \chi_y(x)=e^{jyx},
$$
is an additive ${\bf G}$-valued character:
$$
\chi_y(x_1+x_2)=\chi_y(x_1)\chi_y(x_2), x_1, x_2 \in {\bf R},
$$
$$
\vert \chi_y(x)\vert =1.
$$
We shall use these ${\bf G}$-valued characters on
${\bf R}$ to define an analogue of the Fourier transform
and pseudo-differential operators. We demonstrate that,
besides the ordinary quantum mechanics based on
${\bf C}$-valued characters, there exists another
natural quantum model based on  ${\bf G}$-valued characters --
hyperbolic quantum mechanics. Both quantum models have the same
classical limit.

We also introduce on ${\bf G}$ the positive norm
$$
\Vert z \Vert= \sqrt{x^2+y^2}.
$$
which will be used in analysis over ${\bf G}.$

\section{Ultra-distributions and Pseudo-differential operators over the hyperbolic algebra.}
We recall that for a function $\varphi:{\bf R} \to {\bf C}$ the Fourier transform is defined by
\[\tilde \varphi (p)=\frac{1}{2\pi h}  \int_{-\infty}^{+\infty} e^{\frac{-i p q}{h}} \varphi(q) dq\] and the inverse Fourier transform given by:
\begin{equation}
\label{INV}
\varphi (q)=\int_{-\infty}^{+\infty} e^{\frac{i p q}{h}} \tilde\varphi(p) dp.
\end{equation}
These formulas are well defined for, e.g., functions $\varphi \in {\cal S},$ where $\cal S$
is the space of Schwartz test functions. A pseudo-differential operator $\hat a$ with the symbol $a(q, p)$ is defined by
\begin{equation}
\label{PDO}
\hat a(\varphi)(q)=\int_{-\infty}^{+\infty} a(q, p) e^{\frac{i q p}{h}} \tilde\varphi(p) dp.
\end{equation}
We would like to use the analogous definitions in the case of functions
$\varphi:{\bf R} \to {\bf G},$ and  $a:{\bf R} \times {\bf R} \to {\bf G}$ by
using instead of additive ${\bf C}$-valued characters $ x \to e^{ix}$ additive ${\bf G}$-valued
characters
$ x \to e^{jx}.$ The only problem is that the latter exponent is not bounded and, e.g., the class of functions ${\cal S}$ cannot be used as the base of the hyperbolic Fourier calculus.
Even if we chose the space $D$ of test functions with compact supports, then, for $\varphi \in {\cal D}$, the inverse Fourier transform (\ref{INV}) is in general not well defined.

One of the ways to proceed in such case is to use the theory of analytic generalized functions,
ultradistributions, cf. [7].
Let us consider the space ${\cal A}({\bf R, {\bf G}})$ of analytic functions:
$$
f(x)=\sum_{n=0}^{\infty} f_n x^n, f_n \in {\bf G},
$$
and $||f||_R=\sum_{n=0}^{\infty}||f_n||R^n < \infty, \forall R > 0.$ The $\bf G$-module ${\cal A}({\bf R, G})$
can be endowed with the topology given by the system of norms $\{||\cdot||_R \}.$ This is a complete metrizable
$\bf G$-module (Frechet module). We denote by the symbol $\cal A^{\prime}({\bf R, G})$ the
space of continuous $\bf G$-linear functionals:
\[\lambda:{\cal A}(\bf R, G) \to {\bf G}.\]
Functions $\varphi \in {\cal A}(\bf R, G)$ are called analytic test functions, functionals
$\varphi \in {\cal A}^\prime(\bf R, G)$ are called ({\bf G}-valued) ultradistributions.
As usual in the theory of distributions, we define the derivative of
$\lambda \in {\cal A}^\prime (\bf R, G)$ by
$(\frac{d\lambda}{dx}, \varphi) = -(\lambda, \frac{d\varphi}{dx}).$ This operation
is well defined in the space ${\cal A}^\prime (\bf R, G).$
The Fourier transform of an ultradistribution $\lambda \in {\cal A}^\prime (\bf R, G)$
is the function \[\hat \lambda(y)\equiv {\cal F}(\lambda)(y)=(\lambda(x), e^{j y x}),\; \;  y \in {\bf R}.\]
Properties of the Fourier transform are collected in the following  proposition and theorem:

{\bf Proposition 1.}
{\it{For any ultradistribution $\lambda \in {\cal A}^\prime (\bf R, G)$ its
Fourier transform is infinitely differentiable. We have:}}
\[\frac{d^n}{d y^n} {\cal F}(\lambda)(y)=j^n {\cal F}(x^n \lambda(y));\]
\[{\cal F} \left(\frac{d^n \lambda}{dx^n}\right)(y)=(-j y)^n {\cal F}(\lambda)(y)\].

We denote the Fourier-image of the space of ultradistributions by the symbol $E(\bf R, G).$

We remark that the Dirac $\delta$-function $\delta (x)$ belongs to $\cal A^\prime(\bf R, G)$ and
as always, we have ${\cal F}(\delta^{(n)})=(-j y)^n.$
Thus, in particular, the space $E(\bf R, G)$ contains all polynomials with coefficients belonging to $\bf G.$
The description of the space $E(\bf R, G)$ is given by the following theorem:

{\bf Theorem 1.} (Paley-Wiener) The Fourier-image $E({\bf R, G})$ is equal to the space
$$
\{\varphi \in {\cal A}({\bf R, G}): \left|\left|\frac{d^n \varphi}{d y^n} (0)\right|\right|\leq C_\varphi R_\varphi^n\} .
$$

Thus the Fourier-image consists of $\bf G$-valued analytic functions which have exponentially growing derivatives.
The proof of this theorem is a rather long and we do not present it here. This prove is similar to the prove
of the analogous theorem in superanalysis, see [7].

To proceed to the theory of $\bf G$-valued pseudo-differential operators, we chose the space of symbols
$a (q, p) \in E (Q \times P, {\bf G}),$ where $Q \times  P = {\bf R}^2$ is the (ordinary) phase space.
We can easily generalize all previous constructions to the multi-dimensional case.

The map ${\cal F}: {\cal A}^\prime ({\bf R, G})\to E({\bf R, G})$ is one-to-one. Thus,
for any $v \in E(\bf R, G),$ there exists the unique ultradistribution
$\lambda \in {\cal A}^\prime ({\bf R, G}): {\cal F}(\lambda)= v.$
We denote this $\lambda$ by the symbol $\tilde v.$
We shall also use (as people do in physics) the symbol of integral to denote the action of
an ultradistribution $\lambda$ to a test function $f:\; (\lambda, f)\equiv \int f(x) \lambda(d x).$
In particular, ${\cal F}(\lambda)(y)\equiv \int e^{j y x} \lambda(d x),$
and, for a symbol $a \in E (Q \times P, {\bf G}),$ we have:
\begin{equation}
\label{FTR}
a(q, p)=\int e^{j(q p_1 + pq_1)} \tilde a(dp_1 dq_1).
\end{equation}
To introduce into the model the Planck parameter $h>0,$ we
modify the definition of the Fourier transform for functions
$\varphi$  from the domain of definition of a pseudo-differential operator:
$$
\varphi(q)=\int e^{\frac{j q p}{h}} \tilde \varphi (dp),
$$
where $\lambda=\tilde \varphi \in {\cal A}^\prime.$ At the same time we preserve the definition
(\ref{FTR}) of the Fourier transform for symbols. We define the pseudo-differential operator $\hat a$
with the symbol $a \in E(Q \times P, {\bf G})$ by the natural generalization of (\ref{INV}):
\begin{equation}
\label{INV1}
\hat a(\varphi)(q)=\int a(q, p) e^{\frac{j p q}{h} }\tilde \varphi (dp).
\end{equation}

We remark that $E(P, {\bf G}) \subset {\cal A}(P, {\bf G}).$ Thus the function
$f(p)\equiv a(q, p) e^{\frac{j p q}{h}} \in {\cal A}(P,{\bf  G})$ for any
$q \in {\bf R}.$ Hence we can apply $\lambda \equiv \tilde \varphi \in {\cal A}^\prime (P, {\bf G})$
to the analytic test function $f.$ In principle, the formula (\ref{INV1}) can be used
to define a psedo-differential operator $\hat a$ with a symbol $a \in {\cal A}(Q \times P, {\bf G}).$
However, I do not know how to prove the correspondence principle for this larger class of symbols.

Let $a(q, p)=q.$ Then $\hat a(\varphi)(q)=\int q e^{\frac{j p q}{h}} \tilde \varphi (dp)= q \varphi(q).$

Let $a(q, p)=p.$ Then $\hat a(\varphi)(q)=\int p e^{\frac{j p q}{h}} \tilde \varphi (dp)= \frac{h}{j} \frac{d}{dq} \int e \frac{j p q}{h} \tilde \varphi(dp)=\frac{h}{j} \frac{d}{dq} \varphi (q).$

The first operator $\hat q$  is the position operator and
the second operator $\hat p$ is the momentum operator. This is the hyperbolic Schr\"odinger representation:

$\hat q=q, \hat p=\frac{h}{j} \frac{d}{dq}$

We have the hyperbolic canonical commutation relation

$[\hat q, \hat p]= \hat q \hat p - \hat p \hat q = -hj.$

{\bf Proposition 2.}{\it{Any symbol $a \in E(Q \times P, {\bf G})$ defines the operator}}
\[\hat a: E (Q, {\bf G}) \to E(Q, {\bf G})\]
{\bf Proof.} As always, we define the direct product of distributions $\lambda_1, \lambda_2 \in {\cal A}({\bf R, G}):$
\[(\lambda_1 \otimes \lambda_2 (x_1, x_2), \varphi(x_1, x_2))=(\lambda_1 (x_1), (\lambda_2(x_2),\varphi(x_1, x_2)))\]
for $\varphi \in {\cal A} ({\bf R}^2, {\bf G}).$ This operation ${\cal A}({\bf R, G})\times {\cal A}({\bf R, G})
\to {\cal A} ({\bf R}^2, {\bf G})$ is well defined. We have
$$
\hat a(\varphi)(q)=
\int \left[\int e^{j(p_1 q + q_1 p)} \tilde a(dp_1 dq_1)\right] e^{\frac{j p q}{h}}
\tilde \varphi (dp)=
$$
$$
\int e^{j(p_1 q + q_1 p) + \frac{j p q}{h}} \tilde a \otimes \tilde \varphi (dp_1 dq_1 dp).
$$
Let us consider the ${\bf G}$-linear continuous operator

${\bf S}:{\cal A}(P, {\bf G}) \to {\cal A}( P \times Q \times P, {\bf G}), S(f) (p_1, q_1, p)=f(p+p_1 h) e^{jq_1 p}.$

Then we have
\[\hat a(\varphi)(q)=\int e^{\frac{j q p}{h}} \tilde a \otimes \tilde \varphi \circ {\bf S}(dp).\]
Thus $\hat a(\varphi)={\cal F}(\lambda), \lambda \in {\cal A}^\prime (P, {\bf G}):$
we have
$(\tilde a \otimes \tilde\varphi \circ {\bf S}, f)=(\tilde a \otimes \tilde \varphi, {\bf S}(f)),$
and, since ${\bf S}$ is continuous, $\lambda=\tilde a \otimes \tilde \varphi \circ {\bf S} \in {\cal A}^\prime.$

In fact, any pseudo-differential operator $\hat a: E \to E$ is continuous in a natural topology of inductive
limit on $E.$ However, we shall not use this fact in this paper.

{\bf Proposition 3.} {\it{Any pseudo-differential operator can be represented in the form:}}
\begin{equation}
\label{R1}
\hat a(\varphi)(q)=\int  e^{j q p_1} \varphi(q + h q_1) \tilde a (dp_1 dq_1)
\end{equation}
{\bf Proof.}
We have
$$
\hat a(\varphi)(q)=\int \left[\int e^{\frac{j p(q + hq_1)}{h}}\tilde \varphi(dp)\right] e^{j q p_1} \tilde a (dp_1 dq_1).
$$
{\bf Theorem 1.}
(The formula of composition). {\it For any two pseudo-differential operators $\hat a_1, \hat a_2: E (Q, {\bf G})
\to E(Q, {\bf G})$ with symbols $a_1, a_2 \in E(Q \times P, {\bf G}),$ the composition
$\hat a=\hat a_1 \circ \hat a_2$ is again a pseudo-differential operator with the symbol
$a \in E(Q \times P, {\bf G})$ and}
\begin{equation}
\label{COM}
a(q, p)=a_1 * a_2 (q, p)= \int e^{jq (p_1 + p_2) + jp (q_1 + q_2) + jh q_1 p_2} \tilde a_1 \otimes \tilde a_2 (d p_1 dq_1 dp_2 dq_2)
\end{equation}
{\bf Proof.}By (\ref{R1}) we have:
$$
\hat a_1 (\hat a_2 (\varphi)) (q)= \int e^{j q p_1} a_2 (\varphi) (q + hq_1) \tilde a_1 (d p_1 dq_1)=
$$
$$
\int e^{j q p_1} \left[ \int e^{j(q + hq_1) p_2} \varphi(q + hq_1 + hq_2) \tilde a_2 (d p_2 dq_2)\right]
\tilde a_1 (dp_1 dq_1)=
$$
$$
\int e^{jq (p_1 + p_2)} e^{jh q_1 p_2} \varphi(q + h(q_1 + q_2)) \tilde a_1 \otimes \tilde a_2 (dp_1 dq_1 dp_2 dq_2).
$$
We introduce a ${\bf G}$-linear continuous operator

$B: {\cal A}( P \times Q, {\bf G}) \to {\cal A}( P \times Q \times  P \times Q, {\bf G}),$
$B(f) (p_1, q_1, p_2, q_2)= e^{jh q_1 p_2} f(p_1 + p_2, q_1 + q_2).$

We can write:
\[\hat a_1(\hat a_2(\varphi))(q)=\int e^{jq p_1} \varphi (q + h q_1) \tilde a_1 \otimes \tilde a_2 \circ B (d p_1 dq_1).\]
Since $B$ is a continuous operator, $\lambda=a_1 \otimes a_2 \circ B \in {\cal A}^\prime.$
Thus $\hat a_1 \circ \hat a_2$ is also a pseudo-differential operator and its symbol
$$
a(q, p)={\cal F}(\lambda)(q, p)= \int e^{j(qp_1 + pq_1)} \tilde a_1 \otimes \tilde a_2 \circ B (dp_1 dq_1) =
$$
$$
\int e^{j(q(p_1 + p_2) + p(q_1 + q_2))} e^{j h q_1 p_2} \tilde a_1 \otimes \tilde a_2 (dp_1 dq_1 dp_2 dq_2).
$$
We now introduce on the space $E(Q \times P, {\bf G})$ of symbols the hyperbolic Moyal bracket:
\[\{a_1, a_2\}_* (q, p)= a_1 * a_2 (q, p) - a_2 * a_1 (q, p),\]
where the operation * is defined by (\ref{COM}). We remark that $*=*(h)$ depends on the Planck parameter $h>0.$
Thus the Moyal bracket also depends on $h:\{a_1, a_2\}_{*(h)}.$ On the space of smooth functions $f: Q \times  P \to {\bf G}$ we introduce the
Poisson bracket:
\[\{a_1, a_2\} (q, p)= \frac{\partial a_1}{\partial p} (q, p)  \frac{\partial a_2}{\partial q} (q, p) - \frac{\partial a_1}{\partial q} (q, p) \frac{\partial a_2}{\partial p} (q, p).\]
The space $(E(Q \times P, {\bf G}), \{\cdot, \cdot \})$ is a Lie algebra. It contains the Lie-algebra of
classical mechanics, $E(Q \times {\bf P, R})\{\cdot, \cdot \}).$

{\bf Theorem 2.} {\it{Let $a_1, a_2 \in E(Q \times P, {\bf G}).$ Then}}
\begin{equation}
\label{LI}
\lim_{h \to 0} \frac{j}{h} \{a_1, a_2\}_{*(h)}(q, p)=\{a_1, a_2\}(q, p), \; \; (q, p) \in Q \times  P.
\end{equation}
{\bf Proof.} We have $\{a_1, a_2\}_*(q, q)=$
$$
\int e^{jq(p_1 + p_2) + jp(q_1 + q_2)} \left[e^{j h q_1 p_2} - e^{j h q_2 p_1}\right]
\tilde a_1 \otimes \tilde a_2 (dp_1 dq_1 dp_2 dq_2)=
$$
$$
jh \int e^{jq (p_1 + p_2) + jp (q_1 + q_2)} \left[q_1 p_2 - q_2 p_1\right]
\tilde a_1 \otimes \tilde a_2 (dp_1 dq_1 dp_2 dq_2) + 0(h).
$$
We also have $\frac{\partial a_1}{\partial p} (q, p) \frac{\partial a_2}{\partial q} (q, p)=$
$$
\frac{\partial}{\partial p} \int e^{j(q p_2 + pq_1)} \tilde a_1 (dp_1 dq_1) \frac{\partial}{\partial q}
\int e^{j(qp_2 + pq_2)} \tilde a_2 (dp_2 dq_2)=
$$
$$
\int j^2 q_1 p_2 e^{jq(p_1 + p_2) + jp(q_1 + q_2)} \tilde a_1 \otimes \tilde a_2 (dp_1 dq_1 dp_2 dq_2).
$$
Thus we obtain the following hyperbolic Fourier-representation of the Poisson bracket:
$$
\{a_1, a_2\} (q, p)= \int [q_1 p_2 - q_2 p_1] e^{jq (p_1 + p_2) + jp(q_1 + q_2)} \tilde a_1 \otimes \tilde a_2 (dp_1 dq_1 dp_2 dq_2).
$$
Hence we proved (\ref{I}).

A pseudo-differential operator $\hat a$ is called an observable if its symbol takes only real values:
\[a: Q \times  P \to {\bf R}.\]
The algebra of observables can be identified with the algebra of symbols ($E(Q \times {\bf P, R}), *(h)$),
where the *-product is given by (\ref{COM}). By theorem 2 the Lie-algebra of hyperbolic observables
($E(Q \times {\bf P, R})$, $\{\cdot, \cdot\}_{*(h)}),$ where $\{\cdot, \cdot\}_{*(h)}$ is the
hyperbolic Moyal product, is the deformation of the classical Lie algebra
$(E(Q \times {\bf P, R}), \{\cdot, \cdot\}),$ where $\{\cdot, \cdot\}$ is the ordinary Poisson bracket.

{\bf Conclusion.}{\it{ The hyperbolic quantum mechanics in the limit $h \to 0$ coincides with the classical mechanics.}}

\section{Classical limit of the hyperbolic quantum field theory}
The classical limit for quantum systems with an infinite number of degrees of freedom was investigated
(on the mathematical level rigourosness) in [11]. I used the theory of ultradistributions on infinite
dimensional spaces to build the calculus of infinite-dimensional pseudo-differential
operators \footnote{First time infinite-dimensional pseudo-differential operators were introduced on the
mathematical level of rigorousness by O. G. Smolyanov [12].} and introduce the Moyal
deformation of the Poisson bracket on the infinite dimensional case, see [11] for detail.
The same we can do in the hyperbolic case.

Let $X$ be an infinite dimensional real topological vector (locally convex)
space, e.g., the space ${\cal S}({\bf R}^n)$ of Schwartz test functions or the
space ${\cal S}^*({\bf R}^n)$ of Schwartz distributions. Denote by the symbol $Y$ the {\bf R}-dual
space of $X$ -- the space of ${\bf R}$-linear continuous functionals
$y:X \to {\bf R}$). As always, we use the notation $(y, x)=y(x).$
Denote the space (${\bf G}$-module) of analytic functions $f:X\to {\bf G}$ by the symbol ${\cal A}(X, {\bf G})$
and the space (${\bf G}$-module) of continuous ${\bf G}$-linear functionals $\lambda:{\cal A} \to {\bf G}$ by
the symbol ${\cal A}^\prime(X, {\bf G}).$

We choose ${\cal A}(X, {\bf G})$ as the space of (analytic) ${\bf G}$-valued test functions
and ${\cal A}^\prime(X, {\bf G})$ as the space of ${\bf G}$-valued (ultra)
distributions. ${\bf G}$-valued additive characters\footnote{We recall that $|e^{j(y, x)}|=1, x \in X.$}
$(y \in Y)$ on $X,$
\[x \to e^{j(y, x)}\]
belong to the space of ${\bf G}$-valued analytic functions.
We define the Fourier transform of an ultradistribution $\lambda \in {\cal A}^\prime (X, {\bf G})$ by:
\[\hat \lambda(y)={\cal F}(\lambda)(y)=(\lambda, e^{j(y, \cdot)}),\; \; y \in Y\]

This is an analytic function on the dual space $Y=X^*$ (endowed with the strong topology).
We denote the Fourier image of the space ${\cal A}^\prime (X, {\bf G})$ of ultradistributions by
the symbol $E(Y, {\bf G}).$ By using methods developed in [13] we can try to obtain an internal
description of this ${\bf G}$-module, Paley-Wiener theorem. However, this is not a trivial problem.

It is important for us that $E(Y, {\bf G})$ contains cylindrical polynomials (as well
as ``nuclear polynomials'', see [13]).
Under some topological restrictions on $X$ (so called approximation property, see [13]) the Fourier transform
\[{\cal F: A}(X, {\bf G}) \to E (Y, {\bf G})\] is one-to-one map.
We consider such a class of infinite-dimensional spaces,
e.g., $X={\cal S}({\bf R}^n), Y={\cal S}^*({\bf R}^n),$ or vice versa. Thus, for any $v \in E(Y, {\bf G})$
there exists the unique ultradistribution $\tilde v \in {\cal A}^\prime (X, {\bf G})$ such
that
$$
v(y)={\cal F}(\tilde v)(y)\equiv \int e^{j(y, x)} \tilde v(dy)
$$
(as always, we use the symbol of integral to denote the pairing between an ultradistribution and a test function). In the same way as in the finite dimensional case we introduce pseudo-differential operators with symbols $a \in E(Q \times P, {\bf G}).$ Here the infinite-dimensional phase-space is introduced in the following way.

Let $Q$ be an ${\bf R}$-linear locally convex space which is reflexif. Thus dual space $Q^*= P$ of $Q$
(endowed with the strong topology) has the dual space $P^*=Q.$ The space $Q \times  P$ is the phase-space.
We remark that $(Q \times  P)^*=Q^* \times  P^*= P \times Q.$ In the above scheme we put $X= P \times Q$
and $Y=Q \times  P$ and proceed:
$$
\varphi(q)=\int e^{\frac{j(p, q)}{h}} \tilde \varphi (dp), \; \; \varphi \in E (Q, {\bf G}),
$$
$$
a(q, p)=\int e^{j(p_1, q) + j(p, q_1} \tilde a (dp_1 dq_1), a \in E (Q \times P, {\bf G});
$$
$$
\hat a(\varphi)(q)= \int a(q, p) e^{\frac{j(p, q)}{h}} \tilde \varphi (dp).
$$

By analogy with one dimensional case we prove (cf. [11], [7]):

{\bf Theorem 3.} {\it{For any symbol $a \in E(Q \times P, {\bf G})$
the pseudo-differential operator $\hat a: E(Q, {\bf G}) \to E(Q, {\bf G}).$
For $a_1, a_2 \in E(Q \times P, {\bf G}),$ the operator $\hat a=\hat a_1 \circ \hat a_2$ is
again a pseudodifferential operator with the symbol

$a(q, p)=a_1 * a_2 (q, p)=$
$$
\int e^{j(p_1 + p_2, q) + j(p, q_1 + q_2) + jh(p_2, q_1)} \tilde a_1 \otimes \tilde a_2 (dp_1 dq_1 dp_2 q_2).
$$
We have the correspondence principle (\ref{LI}) where $\{\cdot, \cdot\}_{*(h)}$ and $\{\cdot, \cdot\}$ are the Moyal and Poisson brackets, respectively.}}

\bigskip

{\bf Conclusion.}
{\it{The classical limit of the hyperbolic quantum theory with the infinite-number of degrees of freedom coincides with ordinary classical mechanics on the infinite-dimensional phase-space.}}

Thus we have two deformations of classical field theory: complex second quantization and hyperbolic second quantization.

\section{Hyperbolic fermions and hyperbolic supersymmetry}

Let $B$ be an algebra over a field $K$ and $A$ be a ring which is also a $B$-module and let
operations of ring and module are connected in the natural way
(in the same way as in the case of an ordinary algebra are related operations of a ring and a linear space).
Such an algebraic structure $A$ will be called a $B$-algebra.

The standard example which has been used in this paper is some space $A$ of functions $f:{\bf R}^m \to {\bf G}.$
They are  ${\bf G}$-algebras.

Let us consider a supercommutative Banach ${\bf G}$-superalgebra $\Lambda=\Lambda_0 \oplus \Lambda_1$
(see, e.g., [8]-[10], [7] for the ordinary supercommutative Banach superalgebras over $\bf R$).

For example, $\Lambda$ can be a Grassmann ${\bf G}$-algebra with $n$-generators
$\theta_1, \ldots, \theta_n:$
$$
{\bf G}_n=\{ u =\sum_\alpha c_\alpha \theta^\alpha: c_\alpha \in {\bf G}\}
$$
and $\alpha=(\alpha_1, \ldots, \alpha_n),
\alpha_j=0, 1, \theta^\alpha=\theta_1^{\alpha_1} \ldots \theta_n^{\alpha_n}$ and $ \theta_i \theta_j= -\theta_j \theta_i.$

In the same way as in [7] we should consider ${\bf G}$-superalgebras
$\Lambda=\Lambda_0 \oplus \Lambda_1$ with trivial $\Lambda_1$-annihilators:
$$
^\perp\Lambda_1=\{u\in \Lambda:  u \lambda=0, \forall \lambda \in \Lambda_1\}=\{0\}.
$$

All Grassmann ${\bf G}$-algebras with a finite number of generators have nontrivial
$\Lambda_1$-annihilators. As an example of  a supercommutative Banach ${\bf G}$-superalgebra
with trivial $\Lambda_1$-annihilator we can consider
an infinite dimensional Banach-Grassmann ${\bf G}$-superalgebra, see [7].

We consider the superspace over ${\bf G}: {\bf R}^{k, l}=\Lambda_0^k \times \Lambda_1^l$
and construct the hyperbolic calculus of super pseudo-differential operators by combining
results of section 2 and [7]. We obtain the following result:

{\bf Theorem 4}
{\it{Hyperbolic Moyal super bracket is a deformation of the ordinary Poisson bracket on the superspace.}}

I would like to thank A. Aspect, L. Accardi, L. Ballentine,
D. Greenberger, S. Gudder, G. `t Hooft,  A. Leggett,  V. S. Vladimirov,
I.V. Volovich for fruitful discussions on contextual approach to QM and its hyperbolic
generalizations.

{\bf References.}

1.  A. Yu. Khrennikov, Linear representations of probabilistic transformations
induced by context transitions. {\it J. Phys.A: Math. Gen.,} {\bf 34}, 9965-9981 (2001).
http://xxx.lanl.gov/abs/quant-ph/0105059.

 A. Yu. Khrennikov, Interference of probabilities and number field structure of quantum models.
{\it Annalen  der Physik,} {\bf 12}, 575-585 (2003).

2. J. Hucks, {\it J. Math. Phys.}, {\bf 34}, 5986 (1993).

3. I. Porteous, {\it Clifford algebras and the classical groups}, Cambridge Univ. Press, Cambridge, 1995.

4. S. Ulrych, {\it Phys. Letters} B, {\bf 612}, 89-91 (2005).

S. Ulrych, {\it Phys. Letters} B, {\bf 618}, 233-236 (2005)

S. Ulrych, Relativistic quantum physics with hyperbolic numbers. Preprint (Z\"urich, Switzerland).

5. G. Kustatter, J. W. Moffat, J. Malzan, {\it J. Math. Phys.}, {\bf 24}, 886 (1983).

6. J. E. Moyal, Quantum mechanics as a statistical theory. {\it Proc. Camb. Phil. Soc},
{\bf 45}, 99-124 (1949).

7. G. Dito and D. Sternheimer, Deformation quantization: genesis, developments and metamorphoses.
Deformation quantization (Strasbourg, 2001),  9--54, IRMA Lect. Math. Theor. Phys., 1,
de Gruyter, Berlin, 2002.

8. A. Yu. Khrennikov, {\it Supernalysis.} Nauka, Fizmatlit, Moscow, 1997 (in
Russian). English translation: Kluwer,
Dordreht, 1999.

9. V. S. Vladimirov   and I. V. Volovich,  Superanalysis, 1.
Differential Calculus.    {\it Teor. and Matem. Fiz.}, {\bf 59}, No. 1, 3--27 (1984).

V. S. Vladimirov   and I. V. Volovich ,  Superanalysis, 2.
Integral Calculus. {\it Teor. and Matem. Fiz.}, {\bf 60},
No. 2, 169--198 (1984).

10.  A. Yu. Khrennikov, Functional superanalysis. {\it Uspehi  Matem. Nauk}, {\bf 43}, 87-114 (1988).

11.  A. Yu. Khrennikov, Second quantization and pseudo-differential operators.
{\it Theor. and Math. Phys.,} {\bf 66},  339-349 (1985); The principle of correspondence in quantum
theories of field and relativistics bosonic string. {\it Matematicheskii
Sbornic}, {\bf 180}, 763-786 (1989).

12. O. G. Smolyanov, Infinite-dimensional pseudo-differential operators and Schr\"odinger
quantization. {\it Dokl. Akad. Nauk USSR}, {\bf 263}, 558-561 (1982).

13.  A. Yu. Khrennikov, The Feynman measure on the phase space and symbols of
infinite-dimensional pseudo-differential operators
{\it Math. Notes,} {\bf 37},  734-742 (1985); Infinite-dimensional pseudo-differential operators.
{\it Izvestia Akademii Nauk USSR, ser.Math.}, {\bf 51}, N. 6,
46-68 (1987); A. Yu. Khrennikov, H. Petersson, Theorem of Paley-Wiener for generalized analytic functions
on infinite-dimensional spaces, Ibid,
{\bf 65}, N. 2, 201-224 (2001).

\end{document}